# Wafer-scale arrayed *p-n* junctions based on few-layer epitaxial GaTe

Xiang Yuan[1,2][†], Lei Tang[1,2][†], Peng Wang[3], Zhigang Chen[4], Yichao Zou[4], Xiaofeng Su[5], Cheng Zhang[1,2], Yanwen Liu[1,2], Weiyi Wang[1,2], Cong Liu[3], Fansheng Chen[5], Jin Zou[4,6], Peng Zhou[7], Weida Hu[3]*, Faxian Xiu[1,2]*

[1] State Key Laboratory of Surface Physics and Department of Physics, Fudan University, Shanghai 200433, China

[2] Collaborative Innovation Center of Advanced Microstructures, Fudan University, Shanghai 200433, China

[3] National Laboratory for Infrared Physics, Shanghai Institute of Technical Physics, Chinese Academy of Sciences, Shanghai 200083, China

[4] Materials Engineering, The University of Queensland, Brisbane QLD 4072, Australia

[5] Satellite Remote Sensing Laboratory, Shanghai Institute of Technical Physics, Chinese Academy of Sciences, Shanghai 200083, China

[6] Centre for Microscopy and Microanalysis, The University of Queensland, Brisbane QLD 4072, Australia

[7] State Key Laboratory of ASIC and System, Department of Microelectronics, Fudan University, Shanghai 200433, China

[†] These authors contributed equally to this work.

*Correspondence and requests for materials should be addressed to F. X. and W. H. (E-mail: faxian@fudan.edu.cn; wdhu@mail.sitp.ac.cn)

**Abstract:**

Two-dimensional (2D) materials have attracted substantial attention in electronic and optoelectronic applications with superior advantages of being flexible, transparent and highly tunable.[1-5] Gapless graphene exhibits ultra-broadband[6] and fast[7] photoresponse while the 2D semiconducting $MoS_2$ and GaTe unveil high sensitivity and tunable responsivity to visible light.[8, 9] However, the device yield and the repeatability call for a further improvement of the 2D materials to render large-scale uniformity.[10, 11] Here we report a layer-by-layer growth of wafer-scale GaTe with a high hole mobility of 28.4 $cm^2$/Vs by molecular beam epitaxy. The arrayed *p-n* junctions were developed by growing few-layer GaTe directly on three-inch Si wafers. The resultant diodes reveal good rectifying characteristics and a high photovoltaic external quantum efficiency up to 62% at 4.8 μW under zero bias. The photocurrent reaches saturation fast enough to capture a time constant of 22 μs and shows no sign of device degradation after 1.37 million cycles of operation. Most strikingly, such high performance has been achieved across the entire wafer, making the volume production of devices accessible. Finally, several photo-images were acquired by the GaTe/Si photodiodes with a reasonable contrast and spatial resolution, demonstrating the potential of integrating the 2D materials with the silicon technology for novel optoelectronic devices.

Graphene has attracted exceptional attention in recent years owing to its high carrier mobility,[12] broadband photon absorption,[13] low dissipation rate[14] and compatibility to silicon processes,[15] thus showing a great potential in electronic and optoelectronic applications, such as ultrafast[16, 17] and broadband[6] photodetectors[18]. However, the gapless band structure of graphene significantly hinders the effective modulation of channel resistance,[3] making the high-performance logical devices inaccessible. On the other hand, two-dimensional (2D) semiconductors, such as $MoS_2$ and GaTe, exhibit intriguing prospects in field-effect transistors[19-21] and *p-n* junctions[22, 23], largely attributed to their direct bandgap,[24] high-absorption coefficient and electrical tunability[25, 26]. Indeed, the *p-n* junctions made from electrostatic doping[27] and layer stacking[28-35] have been recently realized for applications in solar cells[36], photodetectors[37], light-emitting diodes[27, 38], and photocatalyst[39]. In particular, remarkable progress has been made on $MoS_2$ photodetectors: $MoS_2$ photoconductors show a high responsivity of 880 A/W;[40] chemically-doped $MoS_2$ *p-n* junctions exhibit a good detectivity of $10^{10}$ at 1.5 V;[41] combined with graphene[42] and $WSe_2$[43], $MoS_2$ vertical heterojunctions demonstrate external quantum efficiency (EQE) of 55% and 34%, respectively. In parallel with these efforts, the emerging 2D GaTe photoconductors with good photosensitivity were reported recently,[9] suggesting another appealing candidate of the 2D materials for optoelectronic applications. However, to date, the GaTe photodetectors involving hetero *p-n* junctions for faster photoresponse in the micron second range have not been demonstrated yet.

To date, most of the 2D semiconductors are prepared by mechanical exfoliation and chemical vapor deposition (CVD). Due to their small lateral scale, they require a dedicated fabrication process, thus presenting a challenge for the mass production of the integrated devices. Molecular beam epitaxy (MBE), however, can be potentially employed for the growth of 2D heterostructures with the capability of choosing the desired functional layers and developing the large-scale, uniform and reliable epitaxial films.

Here we report a layer-by-layer epitaxial growth of wafer-scale 2D GaTe thin films using MBE and the fabrication of arrayed photodiodes devices. The *in-situ* reflection high energy electron diffraction (RHEED) provides a precise control on the film thickness. The as-grown few-layer GaTe possesses a *p*-type conductivity with a carrier concentration of $2.5 \times 10^{18}$ cm$^{-3}$ and a high mobility of 28.4 cm$^2$/Vs at room temperature. The vertical *p-n* junctions formed by directly depositing GaTe on the silicon wafers construct the broadband photodiodes that respond to the entire visible spectrum, thus enabling the image acquisition of real objects. At the wavelength of 632.8 nm, the diodes show a clear photovoltaic effect with EQE up to 62% under zero bias. It is worthwhile to mention that for photodiodes the EQE is typically less than 100%. The response time of 22 μs to the laser illumination is also fast compared with other 2D semiconductors devices. Combined with the endurable 1.37 million operations, our heterojunction photodiodes show fast, stable and robust photoresponse.

To perform the electrical and optical characterizations of GaTe, mica has been chosen as the suitable substrate because of its electrical insulation and high transparency. The atomically flat surface also makes it an ideal substrate for the Van der Waals epitaxial growth[44]. In this study, 2D GaTe thin films with several layers in thickness were directly grown on the freshly cleaved mica substrates layer-by-layer, monitored by *in-situ* RHEED. Here, the term 'layer' refers to one-unit-cell or a so-called quadruple layer consisting of Te-Ga-Ga-Te.[45] Figure 1a is a plot of *in-situ* RHEED oscillations taken from GaTe, indicative of inheriting the 2D growth mode in the GaTe system. The growth time for each quadruple layer is 2.33 minutes thus enabling fine control of thickness. The inset of Figure 1a shows a sharp and streaky RHEED pattern of an 8 nm GaTe thin film, indicative of a high crystallinity and atomically flat surface. Consistent with the RHEED observations, atomic force microscopy (AFM) reveals further evidence of the terrace structures on the growth front with a step of 0.8 nm, corresponding to the single-layer thickness of GaTe.[45] (Fig. 1b) The root mean square (RMS) for each terrace poses a small value of 1.5 Å which reaches the ultimate resolution of our AFM.

To study the structural properties of the GaTe thin films, X-ray diffraction (XRD) and high-resolution transmission electron microscopy (HRTEM) were carried out at room temperature. Figure 1c displays a XRD spectrum, in which the periodic diffraction peaks are due to the mica substrate and the marked diffraction peak can be indexed as (400) atomic plane of a monoclinic layered structure with $C2/m$ space-group symmetry (point group $C_{2h}$).[46] Energy dispersive X-ray spectrum (EDX) for a 10-layer GaTe film (8 nm) was carried out and the atomic ratio of Ga/Te was identified to be one, consistent with the determined structure, with a standard deviation smaller than 5% across the entire wafer. For the HRTEM experiments, GaTe was peeled off from the mica substrate and transferred onto a copper grid (Figure 1d inset). A typical HRTEM image showed in Figure 1d reveals a perfect crystalline quality.

Raman spectroscopy was employed to understand the phonon vibration modes. Figure 1e shows the normalized Raman spectra taken from GaTe thin films with different thickness, in which three dominant peaks can be seen that correspond to the $A_g$ modes in GaTe, identical to the exfoliated ultrathin GaTe sheets [9]. The fact that the peak at 114 cm$^{-1}$ arises with increasing the thickness of GaTe suggests the phonon modes taking place in our GaTe thin films, which are similar to those in the exfoliated GaTe sheets. [9] It is also noted that the distance between the left two peaks strongly depends on the thickness, similar to that in few-layer MoS$_2$.[47] Peak distance increases from 10.0 to 14.5 cm$^{-1}$ when the number of layers decreases from 28 to 2 (Figure 1e). For this reason, the Raman spectroscopy can serve as a sensitive probe to determine the layers of GaTe and verify the film quality, as well as providing evidence of well control of thickness by the MBE technique.

Temperature-dependent Hall measurements were also performed to evaluate the electrical properties of the 8 nm-thick GaTe thin film grown on the mica substrate. The Hall bar devices were fabricated in the centimeter range. Indium was soldered as electrodes and form a perfect Ohmic contract (Figure S1). Surprisingly, the result shows a *p*-type conductivity with a room-temperature hole mobility reaching 28.4 cm$^2$/Vs (Figure S2), the highest mobility reported for the ultrathin GaTe sheets so far. The absorption spectrum measurements (Figure S12) exhibit a 1.65 eV thickness-independent direct bandgap of few-layer GaTe, consistent with our XRD results and previously report theoretical calculations.[9] The unexpected high mobility and the direct bandgap make the GaTe a promising candidate for realizing highly-efficient devices.

For their practical applications, the GaTe thin films were directly deposited on three-inch *n*-dope silicon (100) substrates under the optimized growth conditions to construct vertical *p-n* junctions. To investigate the atomic composition, the EDX spectrum was collected from the cross-section TEM specimen of a 50-layer thick GaTe sample (Figure 2a). The atomic ratio of Ga and Te is deduced to be around one. Note that the detected Cu signals originate from the copper grid. The inset is EDX mapping of Ga and Te element, showing their uniform distribution in the thin films. Hereafter, we will concentrate on the 10-layer thick GaTe for evaluating the device performance unless otherwise mentioned. Also, the results for 4-layer thick GaTe are provided in the supporting information. Figure 2b depicts the Raman spectra of Si, GaTe/mica and GaTe/Si in the upper, middle and bottom panel, respectively. In the GaTe/Si heterojunction, both peaks from Si and GaTe can be well resolved. Importantly, reminiscent of GaTe grown on mica (Figure 1e), the left two prominent peaks are located at the same wavenumbers, indicating the comparable crystalline quality and the film thickness.

To understand its electrical properties, we further conducted electrical measurements to examine the device performance, as schematically illustrated in Figure 2c. Indium was employed as the metal electrodes for both Si and GaTe by simple soldering. A clear rectifying characteristic is observed in current-voltage (*I-V*) measurements (Figure 2d). The rectification ratio, defined as the ratio of the forward and reverse current, reaches ~$10^3$ under 1 V bias at 300 K. The inset of Figure 2d shows a perfect Ohmic contact on GaTe (the red curve) while on silicon there exists a negligible barrier that does not significantly affect the rectifying characteristic. These devices can also work at low temperatures with 20 times higher rectification ratio as the temperature drops to 150 K, as shown in Figure S5. To study the device statistics, 30 diodes across the wafer were randomly selected and their *I-V* characteristics were recorded for comparison. Figure 2e illustrates the statistical distribution of the rectification ratio. Overall, 67% of the diodes are distributed in the range of 1000~2000 (rectification ratio) while only 6% shows value lower than 1000. These results suggest a good repeatability among diodes across the wafer.

Numerical simulation was carried out to elucidate the device operating principles. Figure 2f shows the energy-band diagrams of the *p-n* junction under a voltage bias ($V= \pm 1$ V) and equilibrium condition (0 V). At equilibrium, a one-sided abrupt junction or $n^+p$ heterojunction is formed as the carrier concentration in the heterojunction changes abruptly from electrons of $N_e = 6.38 \times 10^{18}$ cm$^{-3}$ to holes of $N_h = 2.5 \times 10^{18}$ cm$^{-3}$, as shown in the middle panel. Note that nearly 90% of the GaTe epitaxial layer is depleted at the thermal equilibrium condition, which makes the 2D GaTe a good photo-absorption and carrier-generating layer. It is found that the total thickness of the space charge regime expands almost two times under the reverse bias (-1 V) compared to that under the zero bias. The depletion layer entirely disappears under the forward bias of +1 V, in agreement with the *I-V* characteristics (Fig. 2d).

To allow the highest possible absorption of light in the photodiodes, a transparent

50 nm-thick indium-tin oxide (ITO) was deposited on top of GaTe as the anode electrodes (Figure 3a). ITO has a good transmittance in the visible range[48] through which the light can mostly reach the vertical junction. The formation of the Ohmic contact was also confirmed as the prerequisite for high efficiency devices[49] (Figure S6). The lower panel shows the picture of the real sample on which a periodic arrangement of photodiodes can be witnessed (purple squares). Figure 3b presents the negative biased *I-V* characteristics under laser illumination (632 nm focused laser). The reverse current increases dramatically with the increase of the incident power, suggesting the generation of the photocurrent, which is defined by the difference between the illuminated and non-illuminated current ($I_{ph} = I_{light} - I_{dark}$).

The extracted photocurrent increase with the incident power following a power law (Figure 3c), which results from the increased photo-carriers under strong illumination. However, the photo-carrier generation saturates at a high incident power causing the drop of EQE, as shown in Figure 3d. Here, the EQE is defined as the ratio of collected charge carrier to incident photon, *i.e.*, $EQE = \frac{I_{ph}}{q\phi} = \frac{I_{ph}}{q}\frac{hv}{P_{in}}$, where $I_{ph}$ is the photocurrent, $\phi$ is the incident photon flux, $hv$ is the energy of single photon, $q$ is the electron charge and $P_{in}$ is the incident power of light. The measured EQE for our diodes reaches an unexpected high value of 62% at 4.8 μW and zero bias. The dark current at zero bias is smaller than 50 pA (the photocurrent is on the order of μA) indicating no influential residue bias in the system. Normally EQE is smaller than one. Such a high EQE is attributed to the high absorption rate of the device (Fig. S11) and the direct bandgap of GaTe (Fig. S12) which cause highly efficient carrier excitations.[18,]

[50, 51] Other parameters are also extracted for 4.8μW data near zero bias. The responsivity defined as the ratio between photocurrent and input power is 0.32 A/W. For the solar-cell performance, the filling factor, maximum output power and energy conversion efficiency are calculated to be 25%, 0.051μW and 1.1%, respectively. In addition, we observe a lower EQE (40%) from thinner samples. More detailed data and discussions are available in the supplementary information (Fig. S9).

In addition to the superior photo-responsivity and good external quantum efficiency, we conducted the time-resolved photoresponse experiments by periodically turning on and off laser illumination at a frequency of 761 Hz and recording the signal from an oscilloscope. Figure 4a demonstrates a complete on/off circle during the operation in which the photocurrent exhibits a rapid rise and reaches a steady saturation. To quantitatively analyze the response speed on a time frame, we performed the best fit to the rising and falling components, as presented in Figure 4b. The rising edge is perfectly fitted, overlapping with the single exponential function following $I_{ph} = I_{sat}(1 - e^{-(t-t_0)/\tau})$ and yielding a time constant of 22 μs which is relatively fast for the state-of-the-art 2D semiconductors. Although the falling part cannot be completely fitted, the sharp edge indicates that the response time should be much less than 22 μs (the inset of Figure 4b). Furthermore, the signals recorded by the oscilloscope remain nearly unchanged after 1.37 million cycles of operation (Figure 4c), pointing to the excellent stability and reliability of the photodiodes.

To further probe the photoresponse characteristics of the devices, we performed spatially-resolved photocurrent measurements by incorporating a galvanometer into the light path (Figure 4d). The generated photocurrent is uniformly distributed in the junction area (a corner of the device) and the spatial mapping exactly matches the device geometry as shown in the inset of Figure 4a. The statistics of the device as to the photoresponse performance were also analyzed similar to what has been done for the electrical test (Figure 2e). Incident power is increased to 380 µW to make the laser spot easily visible. The photocurrent is mainly distributed in the range of 20 to 40 µA, occupying 73% of 30 randomly-chosen photodiodes (Figure 4e). Only 6% of samples shows a photocurrent below 20 µA revealing a good repeatability. Wavelength-dependent photoresponse measurements were also performed in the visible range under an incident power of 5 µW. Inspiringly, the photoresponse of the diodes covers nearly the entire range of 400~750 nm (Figure 4f). The absorption spectrum (Fig S13) of pure GaTe shows a distinct photovoltaic contribution from the 2D GaTe layers particularly for the wavelength less than 650 nm. On the other hand, for the wavelength higher than 650 nm, GaTe cannot absorb the light efficiently. However, single crystal silicon exhibits every high EQE at this region. Thus, the GaTe and silicon compensate each other, covering the photoresponse of the entire visible range. Moreover, the steady EQE (>60%) in the entire visible spectrum makes the GaTe/Si diodes applicable for broad-band detection or solar cells with high efficiency.

Although the arrayed GaTe/Si *p-n* junctions showed a super-fast, broadband and repeatable photoresponse, they are still at the early age of proof-of-concept. Significant progress must be made towards the practical utilization of the 2D materials. With this in mind, we replaced the CCD unit in a digital camera with our GaTe/Si photodiodes that were connected to a pre-amplifier. Then the camera with GaTe/Si sensors was placed on a piezoelectric platform, as showed in Figure 5 (left upper panel). By scanning the stage in different directions, the spatial photoresponse can be recorded as a function of 2D coordinates. With this technique, several images of printed objects were taken with a reasonable contrast and spatial resolution: "Fudan & SITP", "Two dimensional material" (in Chinese) and an official logo of Fudan University. With the combination of the 2D materials and the traditional silicon technology, it's possible to realize commercial devices upon further improvement on the fabrication process and material synthesis.

In conclusion, we achieved controllable layer-by-layer growth of GaTe by MBE. Few-layer GaTe forms wafer-scale arrayed *p-n* junctions on silicon with high yield and repeatable electrical and optical characteristics. Importantly, they can be used as photodiodes having sensitive, fast, stable and broadband photoresponse. The capability of acquiring real images by the GaTe/Si heterojunctions demonstrates the potential for the 2D materials coming to the practical applications.

**Method**

**Thin film synthesis.**

Few-layer GaTe was grown on fleshly cleaved mica or silicon substrates in a Perkin Elmer 430 molecular beam epitaxy system. Standard Knudsen cells provide high-purity Ga and Te fluxes. Typically, the film growth is under Te-rich conditions with Ga/Te ratio of 1:20 measured by crystal oscillator. The growth temperature was kept at 600℃.

**Characterization tools.**

The surface morphology was examined by AFM (Park System E-120) in a tapping mode. Raman spectroscopy measurements were performed with an incident laser wavelength of 632.8 nm (Renishaw-inVia confocal Raman system). The laser spot is focused with diameter of 3 μm. Structural properties were investigated by using HRTEM (FEI Tecnai F20) and XRD (Bruker D8 Discover). Temperature-dependent Hall measurements were conducted in a Quantum Design physical property measurement system (PPMS) with a maximum magnetic field of 9 T.

**Electrical and optoelectrical characterizations.**

Electrical characteristics and photoresponse measurements were carried out using an Agilent 2902. The voltage accuracy is 0.1 μV. The incident light from a 632.8 nm-laser was focused on the samples. The intensity was controlled by a series of neutral density filters plus a linear polarizer/wave-plate and calibrated by Thorlab-S130C. For the time-resolved photocurrent measurements, the laser illumination was controlled by a Thorlab-ITC4001 controller, working at chopping frequency of 761 Hz. A Tektronix MDO 3014 oscilloscope was used to read the signals. For the spatially-resolved photocurrent measurements, a galvanometer was controlled by an open-source controller, thus allowing the laser spot scanning across the device. For the wavelength-resolved measurements, a Horiba spectrometer combined with a halogen lamp was used, covering the visible range.

**Imaging.**

A digital camera was dissembled. The CCD unit is replaced by the GaTe/Si diodes. The signal was amplified by a pre-amplifier (Stanford Research System 560) and acquired by a home-built acquisition system. By contemporary monitoring the controller position and photoelectric signal, the imaging process was achieved.

**Numerical Simulations.**

To investigate the energy-band diagrams in the $n^+p$ heterojunction under different bias conditions, 2D steady-state numerical simulations were performed using the Sentaurus Device, a commercial software package from Synopsys. For the plain drift-diffusion simulation the well-known Poisson equation and continuity equations are used in the calculations. The carrier generation-recombination process consists of Shockley-Read-Hall, Auger, and radiative terms. Additionally, tunneling effects, such as band-to-band and trap-assisted tunneling, are included in the simulation.

# Acknowledgements

This work was supported by the National Young 1000 Talent Plan, Pujiang Talent Plan in Shanghai, National Natural Science Foundation of China (61322407, 11474058, 11322441), and the Chinese National Science Fund for Talent Training in Basic Science (J1103204). Part of the sample fabrication was performed at Fudan Nano-fabrication Laboratory. We acknowledge Yuanbo Zhang, Yizheng Wu ,Zuimin Jiang, Likai Li, Boliang Chen for great assistance during the device fabrication and measurements.

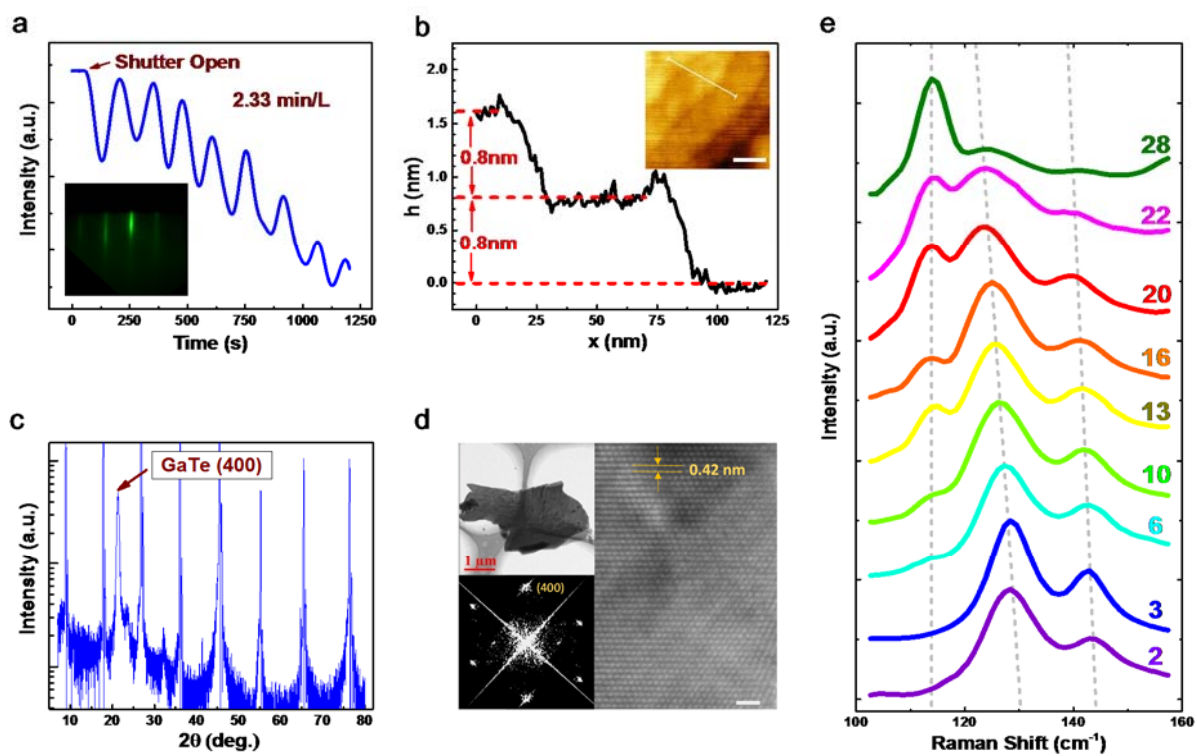

**Figure 1| Growth, structural properties and the thickness-dependent Raman spectra of the few-layer GaTe thin films. a,** RHEED intensity oscillations. Persistent oscillations serve as strong evidence of layer-by-layer growth. The growth rate is determined to be 2.33 min/layer. The inset is a sharp and streaky RHEED pattern for an 8 nm-thick GaTe film, indicative of a high crystalline quality. **b,** The height profile across different terraces on the sample surface. The step of 0.8 nm corresponds to the thickness of one quadruple layer in GaTe. The inset is the AFM image showing flat terraces with RMS less than 1.5 Å. The scale bar is 50 nm. **c,** A typical XRD pattern of GaTe. The marked peak is identified as GaTe (400) plane while others originate from the mica substrate. **d,** A HRTEM image of a GaTe thin flake on a holey copper grid showing superior crystalline quality. This is a top view of GaTe (400) and the scale bar presents 1 um. The inset is a low magnification TEM picture and fast Fourier transform (FFT) image. **e,** Raman spectra for GaTe with different thickness. The number of layers are labeled. As the film becomes thicker, the peak at 114 cm$^{-1}$ emerges and the distance between the left two peaks decreases.

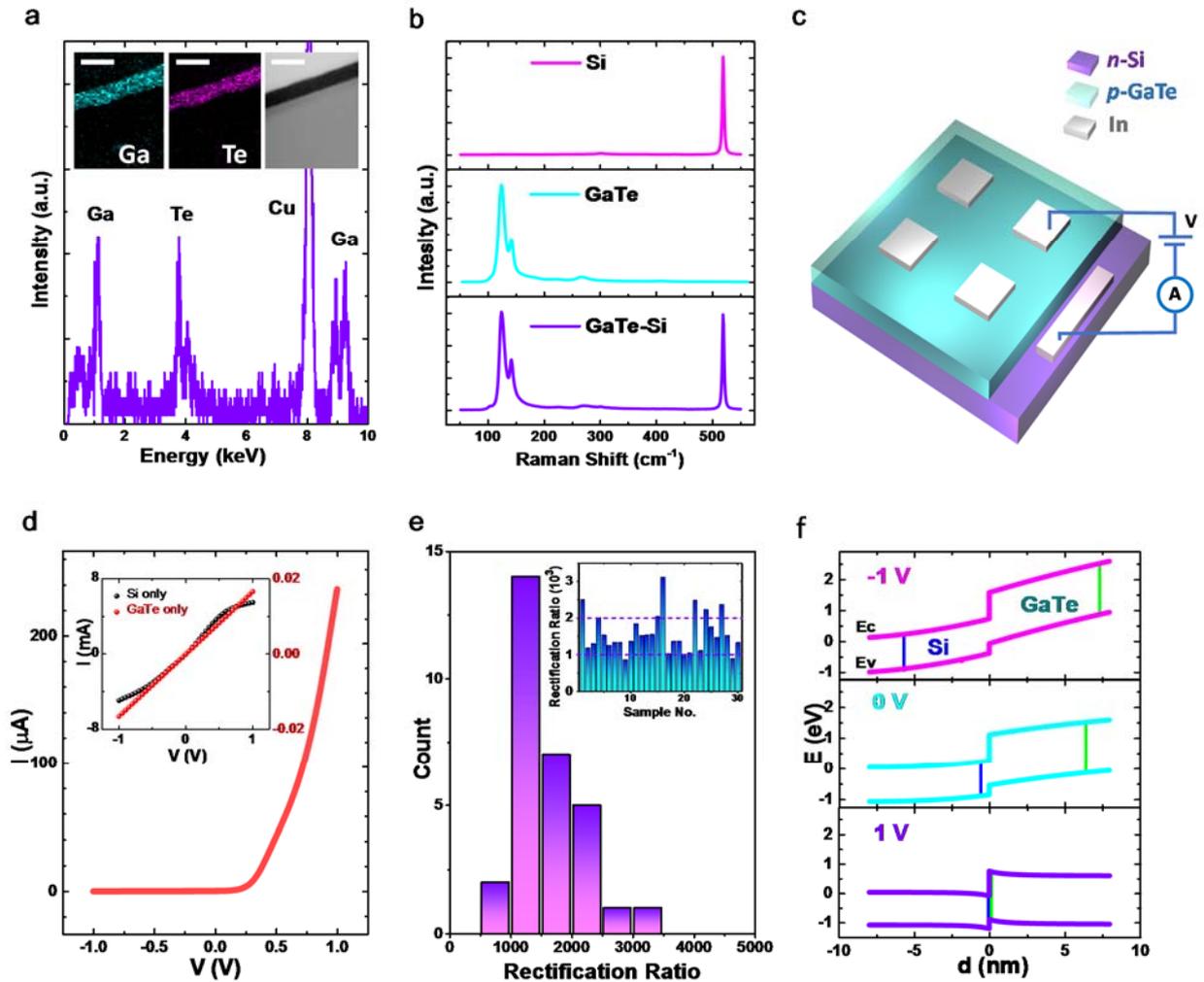

**Figure 2| Electrical characterizations, EDX and Raman spectra of the 10-layer GaTe/Si *p-n* junctions. a,** A typical EDX spectrum taken from the few-layer GaTe grown on Si. The atomic ratio of Ga/Te reaches one within the accuracy of the equipment. The inset is a TEM image and the mapping of the Ga and Te distribution obtained from the EDX scanning in HRTEM, indicating uniform element distribution. The scale bar represents 100 nm. **b,** Raman spectra taken from the area of silicon, GaTe/mica and junction area. Both of the GaTe and silicon peaks can be observed in the junction area. **c,** A schematic view of the device structure. **d,** The rectifying *I-V* characteristics of the device. The inset shows the Ohmic contact for In-GaTe while In-Si presents a negligible barrier. **e,** A histogram of the rectification ratio distribution based on 30 randomly-chosen diodes across a three-inch GaTe/Si wafer. The inset shows the rectification ratio for individual diode. **f,** Calculated band diagrams of GaTe/Si junction under different bias conditions. The space charge regime (between two vertical lines) can be significantly decreased with positive bias which explains the high rectification ratio. GaTe is almost depleted at zero and negative bias.

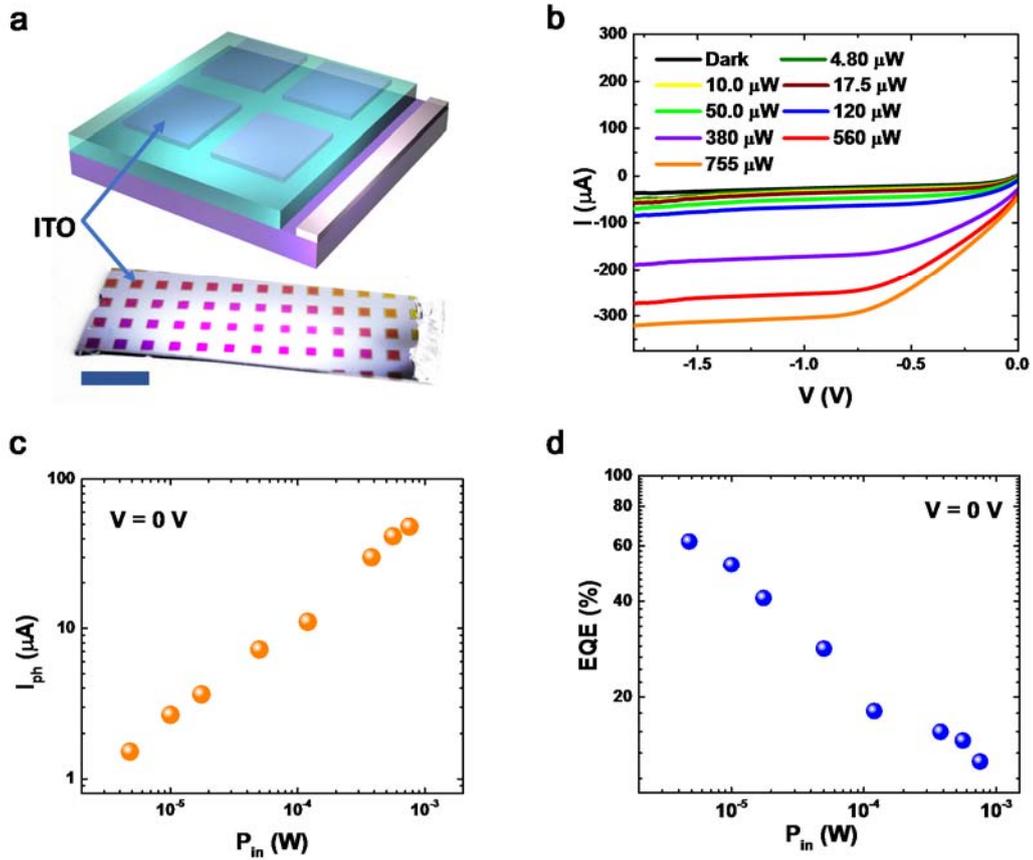

**Figure 3| Photoresponse of 10-layer GaTe/Si *p-n* junctions. a,** A schematic view of the device structure. It shares the same legends with Figure 2**c**. The lower panel shows the picture of the real sample on which a periodic arrangement of devices can be witnessed (colored area). The bar represents 5 mm. **b,** Negative biased *I-V* characteristics under the 632.8 nm laser illumination. When the laser intensity increases, the current systemically rises up to 322 μA, suggesting a high photosensitivity. **c,** Photocurrent under different incident power. **d**, EQE versus incident power at zero bias. A relatively high EQE value of 62% can be achieved at 4.8 μW.

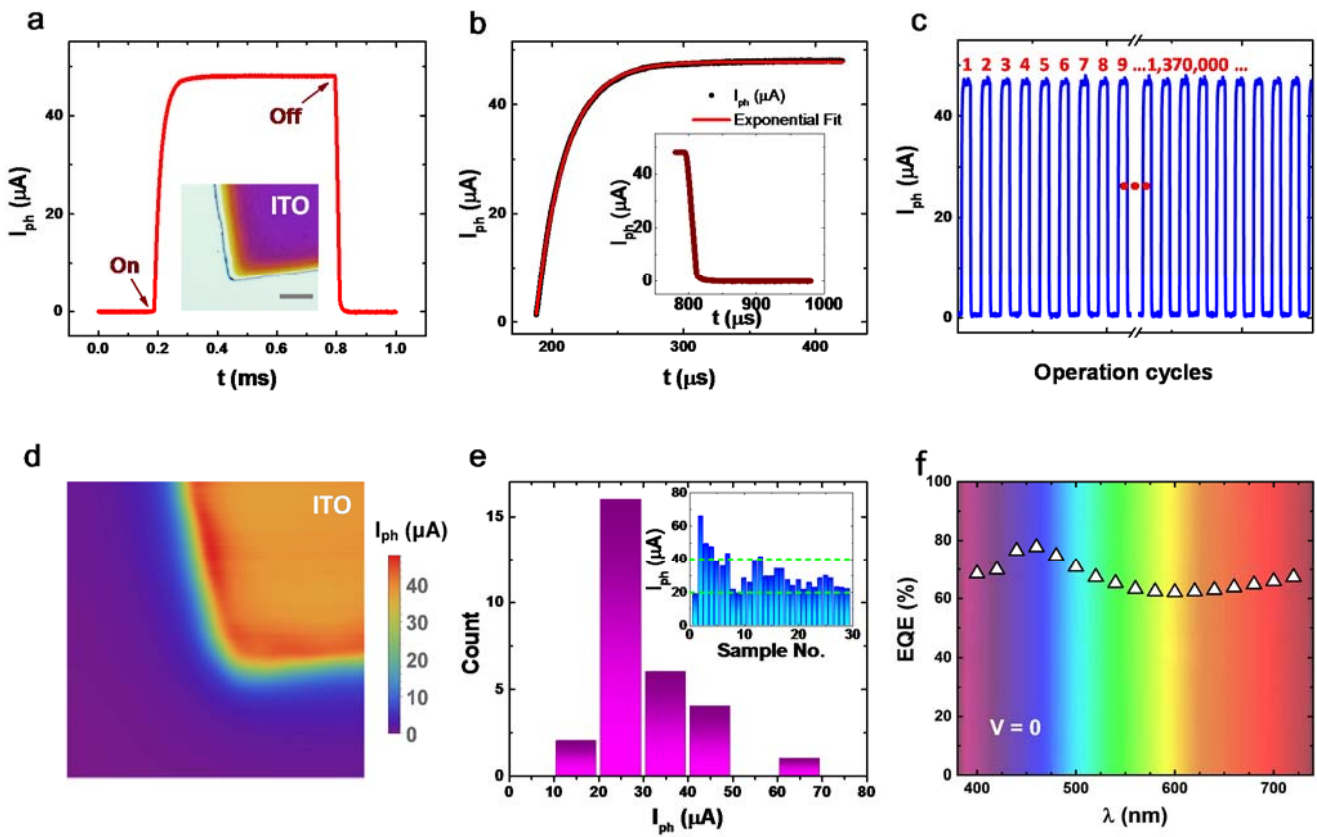

**Figure 4| Time, spatially and wavelength-resolved photoresponse measurements and device repeatability. a,** Time-resolved photoresponse with incident laser power of 4.8 μW. The inset is a microscopic picture from a corner of the device. The bar represents 0.1 mm. **b,** Time-resolved photoresponse. When illuminated, photocurrent rises quickly to reach saturation within a time constant of 22 μs. The black and red curves are the experimental data and exponential fit, respectively. The inset is the falling part of the photocurrent, showing a faster decay with a time constant less than 22 μs. **c,** Photocurrent response during 1.37 million cycles of operation. The device shows an endurable photoresponse. **d,** A spatial mapping of the photocurrent from a corner of the diode. The generated photocurrent is uniformly distributed in the junction area. The spatial mapping exactly matches the device geometry as shown in figure **a** inset. **e,** A histogram of the photocurrent distribution based on 30 randomly-chosen photodiodes. The inset shows the photocurrent for individual diode. They were measured at zero bias under incident power of 380 μW. **f,** Wavelength-dependent EQE at zero bias under incident power of 5 μW. The color of the background is consistent with the wavelength. The GaTe/Si junctions show a uniform and high EQE in the visible range.

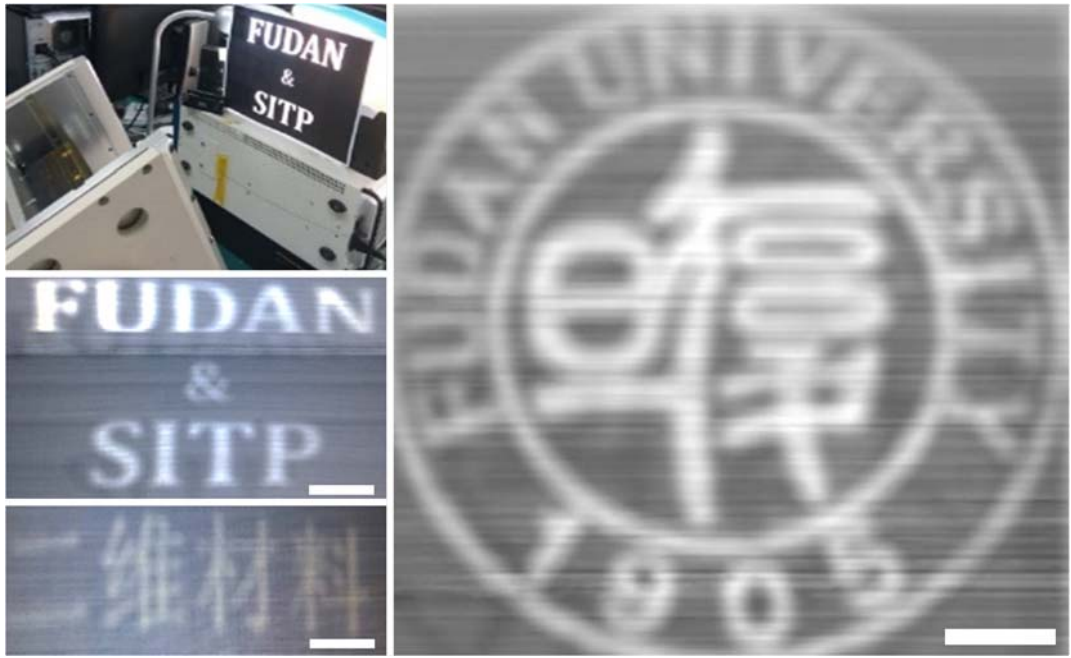

**Figure 5| Experimental set-up and image acquisition by GaTe/Si photodiodes.** Left upper panel shows the experimental set-up, including the piezoelectric platform, camera and the target. All other panels are the images taken by the GaTe/Si diodes with reasonable contrast and spatial resolution: "Fudan & SITP", "Two dimensional material" (in Chinese) and an official logo of Fudan University. The scale bar is 50 mm.

## Electronic Supplementary Material:

Supplementary material is available in the online version of this article.